# Prediction of surface reconstructions using MAGUS

Yu Han[1], Junjie Wang[1], Chi Ding[1], Hao Gao[1], Shuning Pan[1], Qiuhan Jia[1], and Jian Sun[1,*]

[1]*National Laboratory of Solid State Microstructures,*

*School of Physics and Collaborative Innovation Center of Advanced Microstructures, Nanjing*

*University, Nanjing 210093, China*

**Abstract**

In this paper, we present a new module to predict the potential surface reconstruction configurations of given surface structures in the framework of our machine learning and graph theory assisted universal structure searcher (MAGUS). In addition to random structures generated with specific lattice symmetry, we made full use of bulk materials to obtain a better distribution of population energy, namely, randomly appending atoms to a surface cleaved from bulk structures or moving/removing some of the atoms on the surface, which is inspired by natural surface reconstruction processes. In addition, we borrowed ideas from cluster predictions to spread structures better between different compositions, considering that surface models of different atom numbers usually have some building blocks in common. To validate this newly developed module, we tested it with studies on the surface reconstructions of Si (100), Si (111) and 4$H$-SiC (1$\bar{1}$02)-c(2×2), respectively. We successfully gave the known ground states as well as a new SiC surface model in an extremely Si-rich environment.

[*] Corresponding author. jiansun@nju.edu.cn

**Introduction**

Surfaces, as the outermost layer of atoms of solids, usually determine how materials interact with their surroundings. Therefore, surfaces have significant influences on the actual behaviors and properties of materials, such as friction, adhesion, electric or thermal conductivity[1-3], abrasion resistance[4], optical transmission or reflection[4], corrosion resistance[5], catalysis[6-8], biocompatibility, *etc*. The atomic structures of surfaces sometimes differ from their bulk configurations due to the lack of periodic boundary conditions in one dimension, which is known as surface reconstruction. Studying reconstruction behaviors is important for the microscopic understanding of surface properties. For instance, the reconstruction process of catalyst surfaces and different catalytic sites could have great impacts on catalytic performance[7, 8], and the reconstruction behavior of nanowire surfaces, such as reducing dangling bonds, could affect their stability and conductivity[1-3].

Many systems have been identified to exhibit such reconstruction phenomena, including $TiO_2$[9], $RuO_2$[10], C[11], *etc*. It is necessary and urgent to determine their reconstruction model to promote industrial applications and basic research. However, to date, only a few of them have been properly figured out[11-14], most of which are obtained through fitting to the experimental data from scanning tunneling microscopy (STM) or low-energy electron diffraction (LEED) measurements[15, 16]. Usually, it is very difficult to obtain atomic resolution level images experimentally. Moreover, it should be even more difficult to resolve the structures of the second or third layer beneath the outermost surfaces. There is no guarantee that these experimental approaches could find the correct structures, and usually many attempts were made before the final models were recognized. For example, the famous Si (111)-$7\times7$ structure was first recognized by LEED in the 1950s, and different models were proposed, including the DAS model and dimer-chain model, until fully confirmed with the DAS model in the 1990s[15-18]. On the other hand, some specific experimental conditions of pressure and temperature are needed for some particular systems, which is beyond the state-of-art techniques. Given these limits of experiments, it is paramount to develop more efficient theoretical methods to accelerate the discovery of possible surface reconstruction models.

The evolutionary algorithm (EA) method, which aims to predict the structure of global minima of a potential energy surface, was first applied to extended crystal systems[19, 20] and soon broadened into other constrained or low-dimensional materials[21]. It has also been introduced[22, 23] to predict reconstruction models without any prior experimental data. Once given the chemical formula, an EA program can automatically determine the most stable structures under different external conditions. Several implications of programs including USPEX[20], CALYPSO[24], GOFEE[25], *etc.,* have been successfully employed to predict the reconstruction surfaces of several semiconductors[14, 26, 27], metal oxides[12, 13] and ionic compounds[28]. For instance, Lu *et al.*[26] predicted several unexpected surface reconstructions featuring self-assembled nanotubes of diamond that are theoretically stable under specific conditions. Merte *et al.*[12] predicted the new model of the $SnO_2$(110)-(4×1) surface structure, which drew attention decades ago and remained unsolved with an EA search[29] assisting with SXRD findings, and their model matches the experimental patterns very well. Kvashnin *et al.*[28] predicted several new unexpected NaCl surface reconstruction models and provided hints on experiments as well. Although the evolutionary algorithm method has made great progress in this field, we must note that its efficiency is limited by two basic problems: the number of local minima grows exponentially with the number of atoms in the system, and the cost of *ab initio* calculations in local optimizations is very large. It is worth noting that the implication of machine learning methods can dramatically improve the efficiency of structure searches by reducing the cost of structure relaxations.[30]

In this manuscript, we expanded our structure prediction method to surface reconstructions via the framework of MAGUS (machine learning and graph theory assisted universal structure searcher)[31-33]. This method has been successfully employed in a variety of systems and has found many novel structures, including new types of helium water (ammonia) compounds with superionic states[34], layered superconducting materials[35], complicated structures with mixed coordinated atoms[32], and high-energy-density materials[36], which proved its powerfulness in various configurations, including periodic three-dimensional, layered[35] and constrained[37] structures. Through the framework of this program, we now introduce our recently developed surface reconstruction prediction module. The special designs and details of our program are described below, followed by implementations on Si (100)/(111) surfaces and a new 4$H$-SiC (1$\bar{1}$02)-c(2×2) modal to show its effectiveness and potential applications.

**Method**

**Surface slab model used in the evolutionary algorithm**

We constructed a surface slab model in the surface structure-searching module. The typical model is composed of four regions, named the bulk region, buffer region, reconstruction region and vacuum region, as shown in Fig. 1(b). In the searching process, atoms in the bulk region are kept fixed, while the reconstruction region is explored by the algorithm. Atoms in the buffer region only move slightly during structure relaxation. Following the procedure outlined in Fig. 1(a), a number of structures are generated randomly and then relaxed and sorted according to their energy fitness. Structures with relatively low energy are more likely to be chosen and implied with crossover and mutation methods to generate new structures to transfer good 'genes' of configurations to the next generation. The generated new structures are then relaxed, sorted and selected to obtain another generation, and this cycle continues until no new structures with lower energy appear in the next several generations.

**Random structure generation**

The generation of random structures (used mostly in the first generation) can be divided into two categories in our program. The first method, similar to a classic EA method[26, 38, 39], generates the reconstruction region according to 17 plane-group or 80 layer-group symmetry depending on user settings, as shown in Fig. 2(a), before we add it to the top of the buffer region (with the bulk region at the bottom). It is generally believed that applying symmetry could bring more structures with lower energy and have better diversity compared to structures generated randomly without symmetry[24, 38, 39]. For comparison, we generated 400 structures of the Si(111)-3×3 surface with and without symmetry respectively, and calculated their energy. The distribution is shown in Fig. 3(a) (the lowest energy of all structures is set to 0 eV). Among all structures, only 2 structures below 0.5 eV were generated without symmetry, while 18 (by plane group) and 8 (by layer group) structures were generated with symmetry, which is consistent with the previous statement.

In this paper, we also applied a second strategy, which could be seen as a mutation of bulk crystals. We move the atom position randomly from the origin position in the bulk phase, as shown in Fig. 2(b), as well as randomly appending atoms or removing some of them, which is inspired by some natural surface reconstruction processes such as the formation of dimers. The range of random walks of atoms and changes in the numbers of atoms also vary with atomic species and are empirically given by users. Since surface reconstruction comes from the surface of a bulk structure, it is natural to take it as a 'parent' and modify it, and the 'child' structures usually have very low energy since the bulk structure is already the ground state of periodic structures. Among the 400 structures of the Si(111)-3×3 surface generated in this way, 67.3% have energies lower than 1 eV above the global minimum. However, this method generates very similar structures, which is less favorable in EAs, so we only use this method in the first several generations to obtain a quick start of an EA run.

**Heredity and mutation strategy**

We employed the following operations to create offspring during the evolutionary process. (a) Crossover operation, or 'cut and splice' operation, which cuts the two 'parents' into two halves each and splices one part into another one of the other parent, as a common method used in EAs[40, 41]. (b) Layer-slip mutation. It could be viewed as a modification of slip mutation introduced by Liu[42], which randomly chooses one direction and slips the atoms whose fractional coordinates are greater than 0.5 by a random distance. In our algorithm searching for surface reconstruction, we slip all the reconstruction regions along a random direction perpendicular to the nonperiodic axis.

(c) Symmetrization mutation. This mutation method aims to get individuals into "higher" symmetry and reduce the rate of *P1* structures in the population. This method first cuts the parent into pieces and then chooses one part and rotates it around an axis parallel to the cutting plane (referred to as "twinning mutation" in[21, 43] or "piece rotation" in [44]) or reflects it through the cutting plane (referred to as 'piece reflection' in [44]). Considering the differences in periodic conditions between clusters and crystals, the cutting planes are parallel to the nonperiodic axis. (d) We also used some other commonly used mutation strategies in EAs, including (I) permutation, which randomly chooses two atoms and exchanges their atom species[20, 29, 42]. (II) rattle mutation, in which a number of atoms are selected and their positions are moved into nearby positions[29, 45]. (III) ripple mutation, which shifts the coordinates of each atom along a randomly chosen axis by a certain amount[40, 42]. (e) Specially, we applied the strategy of removal or addition of one or more atoms from the structure. It is shown that it is useful to spread the good structures between different compositions when searching for clusters[46], and we found that it also performs well when searching for surfaces that have one nonperiodic axis.

Despite examples of the Si DAS model and SiC DP-H3 model discussed in detail in the Results section, we further performed a systematic exploration searching for the $SnO_2(110)$-(4×1) surface[12] with the suggested $Sn_6O_6$, $Sn_6O_4$, and $Sn_6O_8$ composition of the reconstruction region. We have successfully found a ground state ($Sn_6O_6$) similar to what Merte *et al.* proposed, as well as the lowest-energy structure for $Sn_6O_8$ similar to what they mentioned in the Supporting Material. We also found a relatively low energy model for $Sn_6O_4$, as shown in Fig. 3(b). These three structures only differ in the occupation of two more oxygen atoms marked by the green circle. This suggests that good surface structures have similarities between different compositions, which we could use to perform mutations.

**Computation details**

For the Si surface, we used a Gaussian approximation potential (GAP) for silicon by Bartók *et al.*[47] applied with QUIP[48] code for structural relaxation. First-principles local relaxations of SiC and SnO₂ were performed using the Vienna Ab initio Simulation Package (VASP) code[49] with projector augmented-wave potentials and the Perdew *et al*. exchange-correlation functional (PBE)[50]. To reduce the time cost of global optimizations, we chose a thin slab model containing 3-4 atomic layers during the exploration, and we used DFT calculations at a low accuracy level. Next, we selected several energy-lowest structures of the final generation and recalculated their surface energy to obtain a more accurate result. The final structures were expanded to 8 atomic layers with a vacuum space of 20 Å, and the bottom surface was passivated by hydrogen atoms. In the final DFT calculation, the kinetic energy cutoff for the plane wave basis set was set to 550 eV. The Brillouin zone sampling resolution was $2\pi \times 0.05$ Å$^{-1}$, and the tolerance of the atomic force was set to 0.01 eV/Å.

**Fitness function**

In our surface structure prediction module, the formation energy indicates the relative stability of individuals among the population, which is defined as $E_{form} = E_{total} - E_{slab} - \sum_i n_i \mu_i$, where E$_{total}$ and E$_{slab}$ are the total energy of the reconstructed structure and the clean surface, and $n_i$ and $\mu_i$ are the number difference of reconstructed structure from the clean surface and the chemical potential of atom type i in the reconstructed form, respectively. Considering the situation of a binary compound AB$_x$, if we define its Gibbs free energy $E_{compound}$ as $\mu_A + x\mu_B$, then E$_{form}$ can be rewritten as $E_{form} = E_{total} - E_{slab} - n_A E_{compound} - (n_B - xn_A)\mu_B$, in which all the variables are independent. As we know, if atom type A has a higher chemical potential in the reconstructed structure than in its bulk form, it is very likely to condense on the substrate surface. Therefore, the chemical potential should satisfy the following constraints: $\mu_A < \mu_A^{bulk\_A}$ and $\mu_B < \mu_B^{bulk\_B}$. As a result, the independent variants $\mu_B$ can only change from $\frac{1}{x}(E_{compound} - \mu_A^{bulk\_A})$ (B-poor limit) to $\mu_B^{bulk\_B}$ (B-rich limit), corresponding to the decomposition of the compound to mark the boundary conditions. The calculated E$_{form}$ as a function of $\mu_B$ forms a phase diagram (for example, Fig. 5 for SiC structure), in which the structure with the lowest E$_{form}$ is the most stable one. In the following, we applied our module to several surface reconstruction models to test its performance with fixed or variable numbers of atoms.

**Results**

**1. Si (100) and (111) surface**

Silicon surfaces have been one of the most extensively investigated surfaces, and several reconstruction models have been proposed and observed. To test our method, we performed the EA search on a 1×2 reconstruction of the Si(100) and (111) surfaces and then expanded the surface cell size to 3×3 to deal with more surface atoms.

First, we searched for the so-called "dimer" reconstruction of the Si(100)-(1×2) surface, which is well known as a common building block in surface reconstructions of IVth-group elements such as Si, C, and Ge. A dimer originates from two dangling bonds (DBs) on the cleaved surface, which could form an additional bond to lower its surface energy. Despite the energy reduction of DBs, dimers on Si(100) also become asymmetric due to the Jahn-Teller effect[51, 52], as shown in Fig. 4(a). Our program correctly predicted the asymmetric dimer reconstruction of the Si(100)-(1×2) surface, and the symmetric dimer was also predicted as the second lowest energy structure.

Next, we focus on the Si(111) surface. For the 1×2 size, a "π-bonded chain" reconstruction[53] is the most stable mode. They form two slightly different isomers, positive buckling (pb) and negative buckling (nb)[54], as shown in Fig. 4(b), which also have very close energy[55, 56]. Our program correctly predicted the nb structure as the most stable one, followed by the pb structure as the second-best one. The buckling model[57], in which alternate rows of surface atoms are shifted in and out of the surface, is also predicted as a low-energy structure.

Then, we focus on a more challenging model with a larger cell size to test our program when dealing with more atoms. The dimer-adatom-stacking-fault (DAS) reconstruction model with size 3×3 has 70 Si atoms in a primitive cell (4 ML). The DAS model was first proposed by Takayanagi *et al.*[15] for 7×7 reconstruction with complex building blocks, including adatoms on the top layer, dimers on the second layer, a hole at the cell corner and a half stacking fault layer, as shown in Fig. 4(c). We performed a global search of the Si(111)-(3×3) surface with even atom numbers in the range of 68-72 atoms per unit cell and obtained the DAS model as the best structure. We found that among all the structures of 68 atoms, the without-adatom DAS structure is the most stable, and our program also obtained a DAS structure by adding atoms (Mutation strategy e) to it. To further analyze our results, a principal component analysis (PCA) of all structures visited by our evolutionary program is performed, and the structures are represented by the many-body tensor representation (MBTR)[58] descriptor implied in the DScribe library[59], as shown in Fig. 4(e). We used dark color to show the previous generations and gradually changed it to red among all 50 generations. We also marked the three lowest energy structures, including the target DAS model, with blue marks. Our program generated a relatively wide range of structures at first and became more concentrated at lower energy regions (near the lowest energy structures) as the procedure went on.

## 2. 4$H$-SiC (1$\bar{1}$02)-c(2×2) surface

We now focus on the reconstruction models with more than one atom species. We explored the 4$H$-SiC (1$\bar{1}$02)-c(2×2) surface, which could be obtained by a diagonal cut through the 4$H$-SiC bulk unit cell, first proposed by Virojanadara *et al.*[60, 61]. They also found the existence of one Si adatom per primitive surface cell, known from the STM image[60]. This surface structure was later investigated by Baumeier *et al.*[62] by comparing the surface energy of different proposed models and found the most stable structure, containing building blocks of Si adatoms residing in H3 or T4 sites in addition to dimer reconstructions of C-C atoms.

We performed a global search aimed at the ground state of SiC surface models with different numbers of adatoms, ranging from zero to two, of both Si and C atoms. Our program successfully predicted the known DP-H3 model as the ground state, consistent with the conclusion of Baumeier *et al.*[62], along with some other low-energy models at the Si-rich limit and Si-poor limit, respectively, as shown in Fig. 5(c)-(e), and compared their formation energy as a function of the chemical potential of Si atoms, including all the most stable structures of a given composition.

Among all structures predicted by our program, a SiC surface with two C adatoms is the most unstable composition with relatively high formation energy. A clean SiC surface without adatoms forms C-C dimers to reduce its energy with a DP structure, as shown in Fig. 5(c). The DP-H3 model (Fig. 5(d)) has the lowest formation energy at a wide range of Si chemical potentials, which runs from -5.97 eV/atom (Si-poor limit) to -5.53 eV/atom. This once again shows that implying strategy (e) in an evolutionary program could speed up the procedure by adding a Si atom to the DP structure and successfully obtain the correct ground state. We also found that at the Si-rich limit, a DP model with two Si atoms near H3 sites (Fig. 5(e)) is preferred, and this one more Si adatom brings the surface Cm symmetry, while the DP-H3 model takes P1 symmetry. The PCA of a SiC searching run is shown in Fig. 5(b). We found that unlike Si(111) surfaces, the lowest energy structures of different compositions are not very far from each other, marked by cyan crosses, which suggests that they share similar atomic environments, which agrees with our result that some basic surface building blocks such as H3/T3 adatoms and dimers are common in low energy surface models.

**Conclusion**

In summary, we demonstrated a new module to predict surface reconstructions in the framework of machine learning and graph theory aided crystal structure searching (MAGUS) developed in our group. We built a surface slab model to replace the periodic crystal cell and used several new generation methods for generating both random structures and offspring structures. The random walk of atoms in bulk materials was used to simulate some natural reconstruction processes, as well as a specific mutation to spread building blocks of lower energy structures. We applied this new module to explore several Si (100) and (111) surface models to show its power in exploring various types of surface reconstructions and employed it on a 4*H*-SiC ($1\bar{1}02$)-c(2×2) surface to perform a systematic study on surfaces with different compositions. Despite the previously known DP-H3 ground state of the SiC surface, we also found a new model that is more energetically favorable in Si-rich environments. For further applications, our new module is also able to predict adsorption-induced surface reconstructions, which can be employed to study catalytic processes, surface oxidation processes, *etc*.

**Acknowledgment**

J.S. gratefully acknowledges the financial support from the National Natural Science Foundation of China (grant nos. 12125404, 11974162, and 11834006), the National Key R&D Program of China (grant nos. 2022YFA1403201), and the Fundamental Research Funds for the Central Universities. The calculations were carried out using supercomputers at the High Performance Computing Center of Collaborative Innovation Center of Advanced Microstructures, the high-performance supercomputing center of Nanjing University.

## References

[1] Y. He, G. Galli, Microscopic origin of the reduced thermal conductivity of silicon nanowires, Phys Rev Lett, 108 (2012) 215901.

[2] I. Ponomareva, D. Srivastava, M. Menon, Thermal Conductivity in Thin Silicon Nanowires: Phonon Confinement Effect, Nano Letters, 7 (2007) 1155-1159.

[3] T. Vo, A.J. Williamson, G. Galli, First principles simulations of the structural and electronic properties of silicon nanowires, Physical Review B, 74 (2006).

[4] N. Yokoi, K. Manabe, M. Tenjimbayashi, S. Shiratori, Optically Transparent Superhydrophobic Surfaces with Enhanced Mechanical Abrasion Resistance Enabled by Mesh Structure, ACS Applied Materials & Interfaces, 7 (2015) 4809-4816.

[5] L. Ma, F. Wiame, V. Maurice, P. Marcus, Origin of nanoscale heterogeneity in the surface oxide film protecting stainless steel against corrosion, npj Materials Degradation, 3 (2019).

[6] M. Matsukawa, R. Ishikawa, T. Hisatomi, Y. Moriya, N. Shibata, J. Kubota, Y. Ikuhara, K. Domen, Enhancing Photocatalytic Activity of $LaTiO_2N$ by Removal of Surface Reconstruction Layer, Nano Letters, 14 (2014) 1038-1041.

[7] F. Polo-Garzon, Z. Bao, X. Zhang, W. Huang, Z. Wu, Surface Reconstructions of Metal Oxides and the Consequences on Catalytic Chemistry, ACS Catalysis, 9 (2019) 5692-5707.

[8] G.A. Somorjai, Surface Reconstruction and Catalysis, Annual Review of Physical Chemistry, 45 (1994) 721-751.

[9] A.L. Linsebigler, G. Lu, J.T. Yates, Jr., Photocatalysis on $TiO_2$ Surfaces: Principles, Mechanisms, and Selected Results, Chemical Reviews, 95 (1995) 735-758.

[10] H. Over, Y.D. Kim, A.P. Seitsonen, S. Wendt, E. Lundgren, M. Schmid, P. Varga, A. Morgante, G. Ertl, Atomic-Scale Structure and Catalytic Reactivity of the $RuO_2$(110) Surface, Science, 287 (2000) 1474-1476.

[11] S. Lu, D. Fan, C. Chen, Y. Mei, Y. Ma, X. Hu, Ground-state structure of oxidized diamond (100) surface: An electronically nearly surface-free reconstruction, Carbon, 159 (2020) 9-15.

[12] L.R. Merte, M.S. Jorgensen, K. Pussi, J. Gustafson, M. Shipilin, A. Schaefer, C. Zhang, J. Rawle, C. Nicklin, G. Thornton, R. Lindsay, B. Hammer, E. Lundgren, Structure of the $SnO_2$(110)-(4×1) Surface, Phys Rev Lett, 119 (2017) 096102.

[13] H.A. Zakaryan, A.G. Kvashnin, A.R. Oganov, Stable reconstruction of the (110) surface and its role in pseudocapacitance of rutile-like $RuO_2$, Sci Rep, 7 (2017) 10357.

[14] Q. Wang, A.R. Oganov, Q. Zhu, X.F. Zhou, New reconstructions of the (110) surface of rutile $TiO_2$ predicted by an evolutionary method, Phys Rev Lett, 113 (2014) 266101.

[15] K. Takayanagi, Y. Tanishiro, S. Takahashi, M. Takahashi, Structure analysis of Si(111)-7×7 reconstructed surface by transmission electron diffraction, Surface Science, 164 (1985) 367-392.

[16] K. Takayanagi, Y. Tanishiro, Dimer-chain model for the 7× 7 and the 2×8 reconstructed surfaces of reconstructed surfaces of Si(111) and Ge(111), Phys Rev
**12/20**

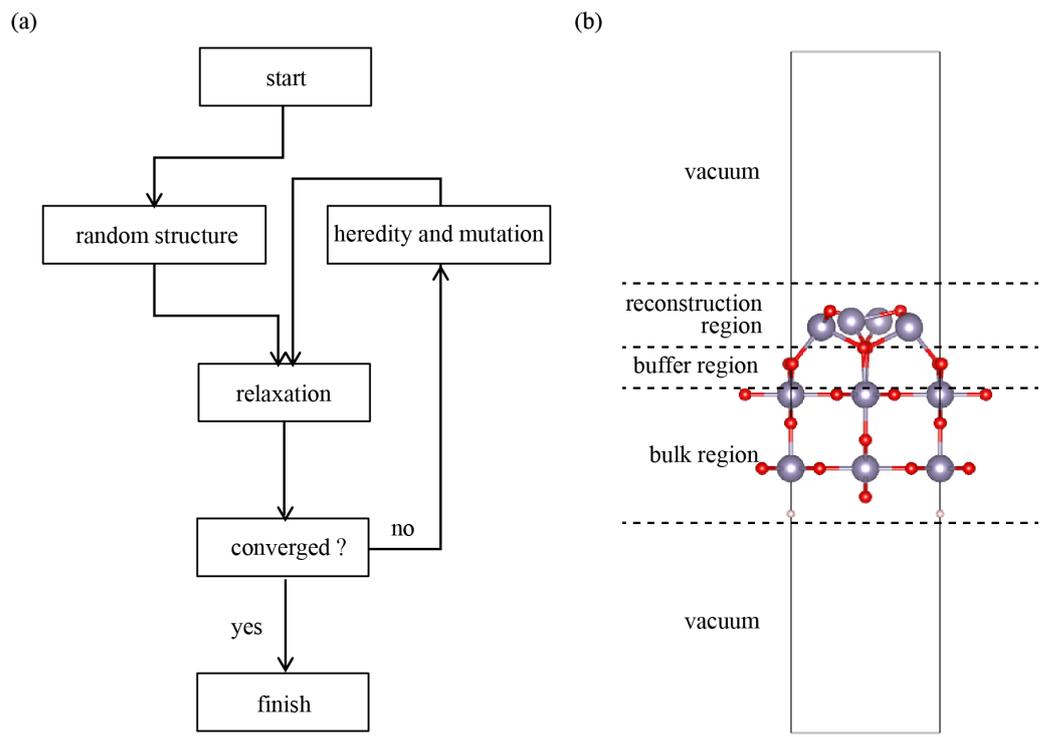

**Fig. 1.** (a) Flowchart and (b) a typical surface slab model used in our surface reconstruction prediction algorithm.

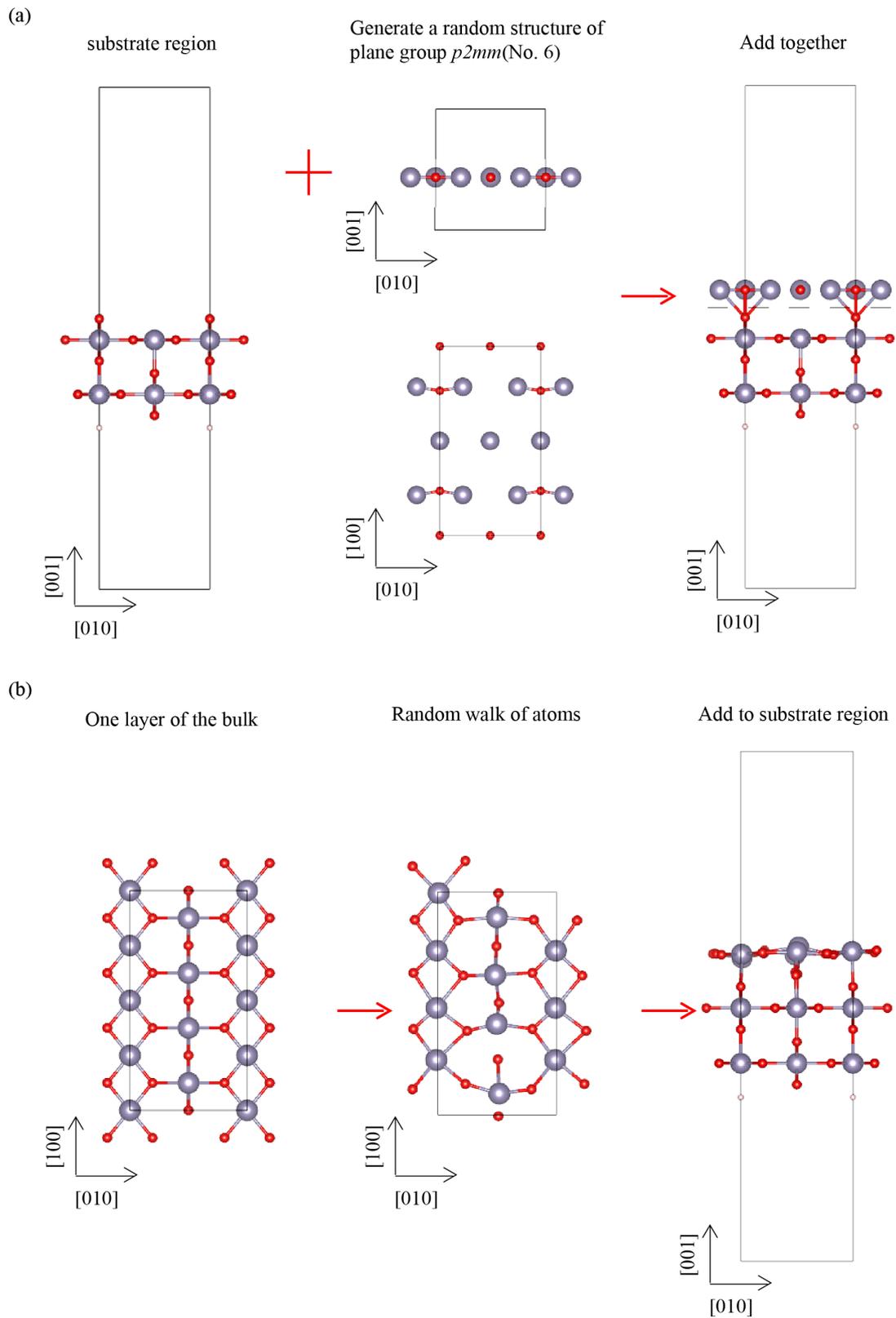

**Fig. 2.** Generation of random structures based on (a) symmetry settings and (b) bulk structure (taking the SnO$_2$(110)-(4×1) surface as an example).

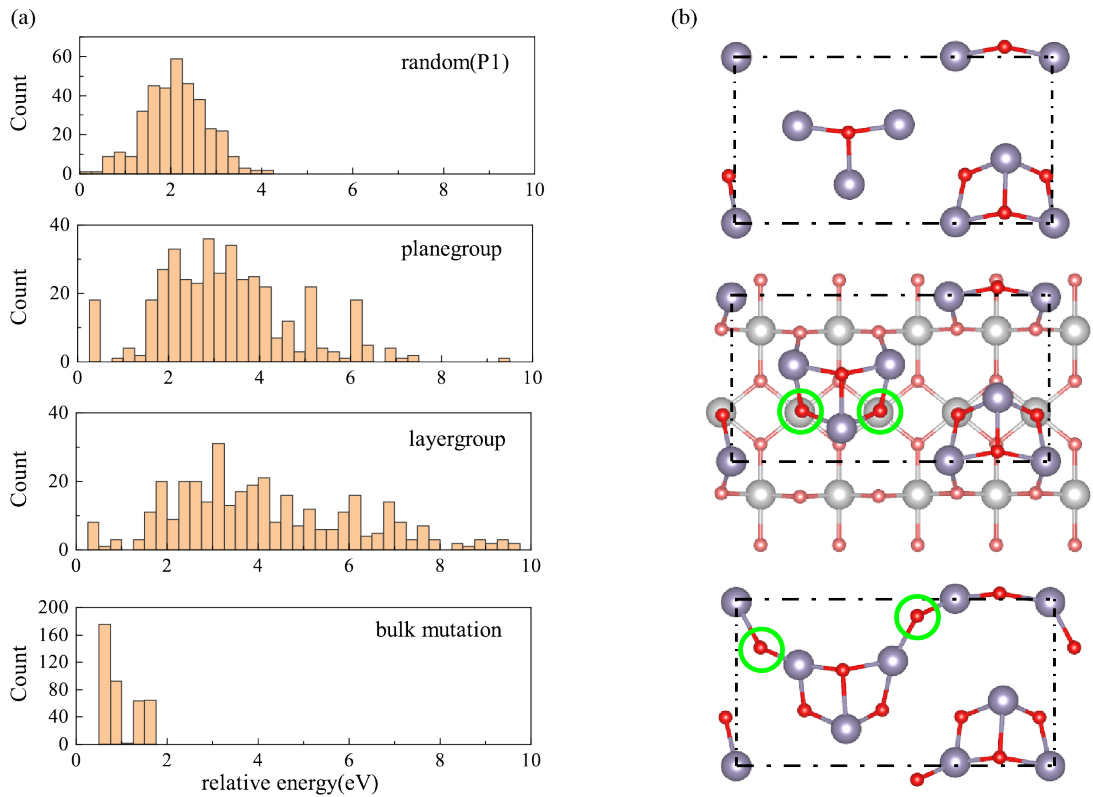

**Fig. 3.** (a) Relative energy of 400 random Si(111)-3×3 surface structures generated without symmetry, with plane-group symmetry, with layer-group symmetry, and based on bulk mutations. The lowest energy of all relaxed structures is set to 0 eV. (b) Metastable $Sn_6O_4$ and lowest energy structure of $Sn_6O_6$ and $Sn_6O_8$ of the $SnO_2$(110)-(4×1) surface structure model of different compositions. Their differences in O occupancies are marked by green circles.

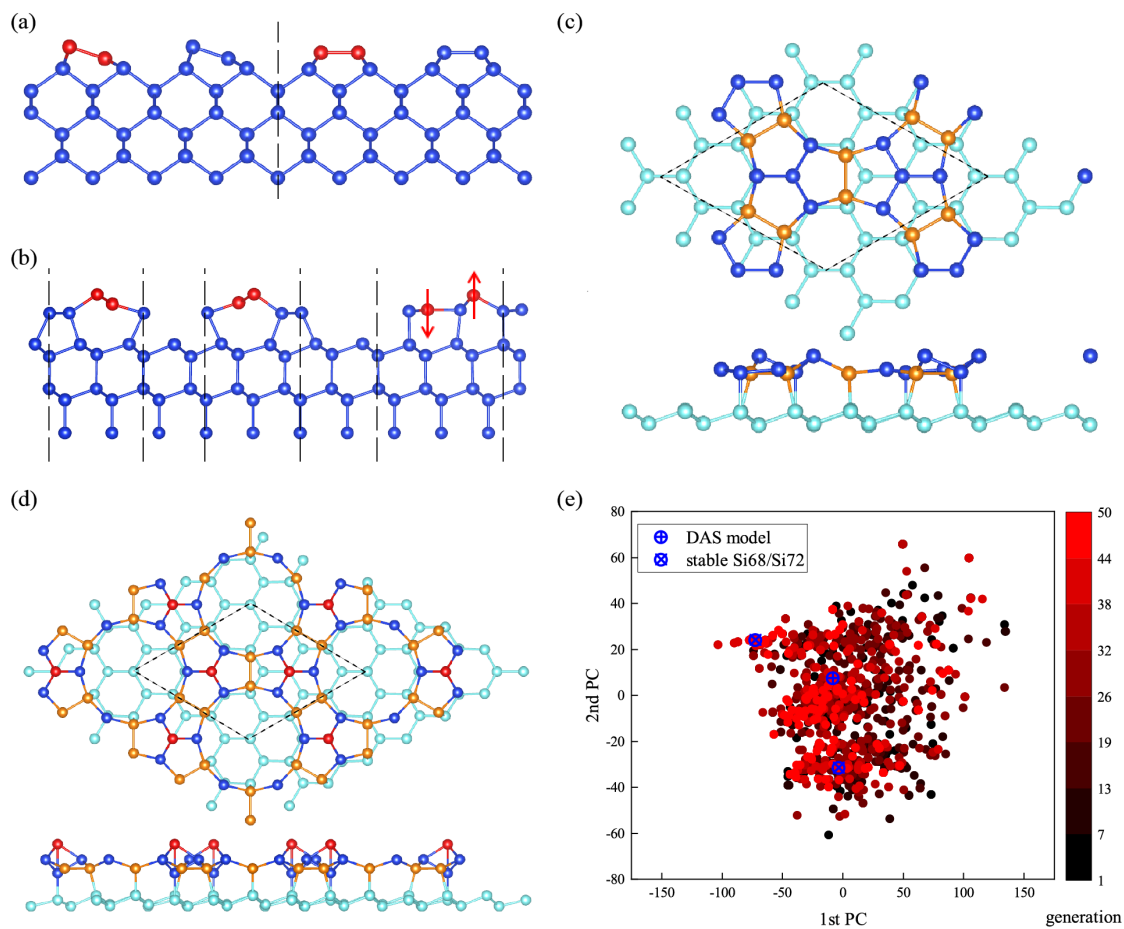

**Fig. 4.** Reconstruction models of the Si surface. (a) Si(100)-(1×2) asymmetric and symmetric dimer. (b) Si(111)-(1×2) negative buckling, positive buckling "π-bonded chain" model and buckling model. (c) Si(111)-(3×3) dimer-adatom-stacking-fault model; dimers are marked by orange and red for adatoms. The faulted half of the unit cell is on the left. (d) Lowest energy model of 68 atoms (4ML) found by our program. It looks almost the same as the DAS model except for the 2 adatoms on the top layer. (e) Structures visited in the EA run. Blue marks represent the most stable structure of 68 atoms, 72 atoms (4ML), and the DAS model, respectively.

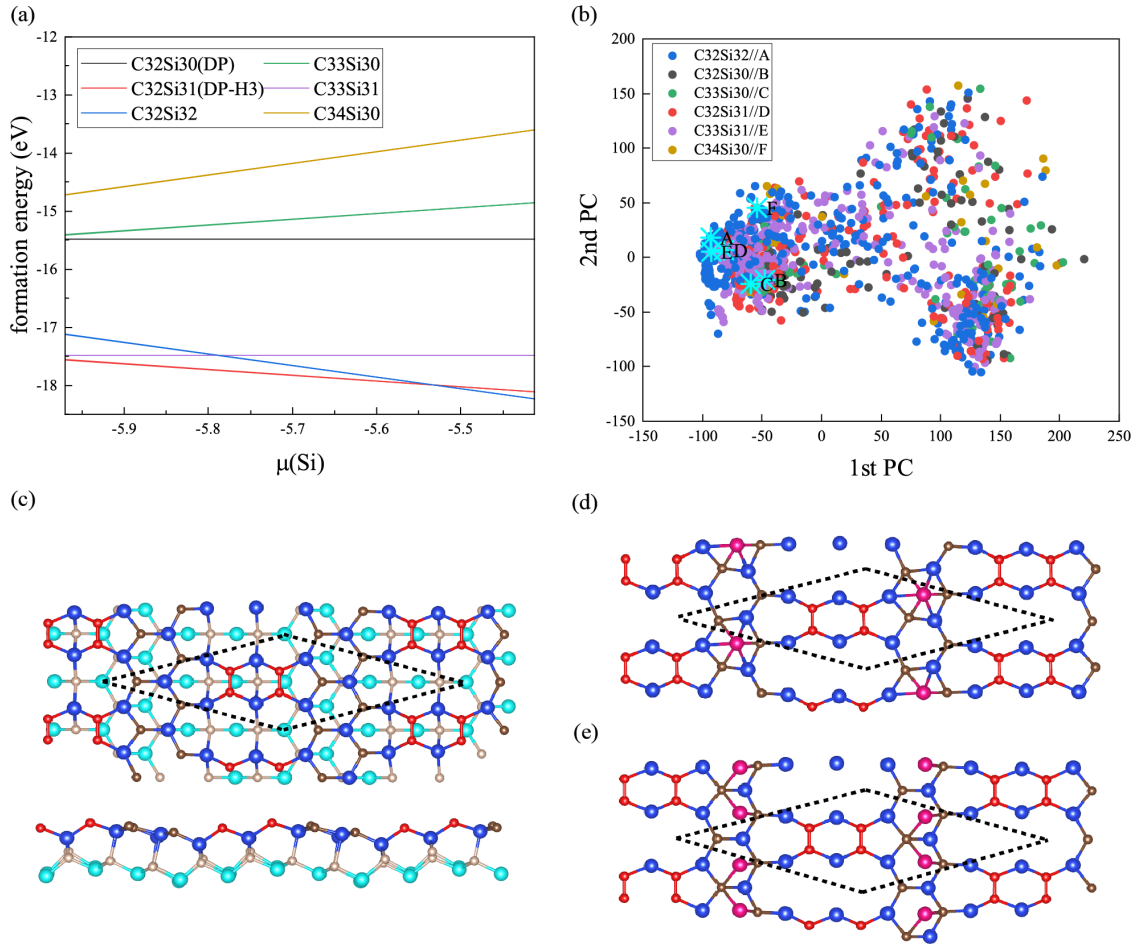

**Fig. 5.** Reconstruction models of the 4H-SiC (11$\bar{0}$2)-c(2×2) surface. (a) Phase diagram with different compositions. (b) Structures visited in the EA run. Cyan crosses represent the lowest energy model of each composition. (c) Dimer pairs (DP) model. Red atoms indicate the C-C dimers. (d) DP-H3 model. One Si adatom (pink) resides at the H3 site. (e) Lowest energy model of $C_{32}Si_{32}$(4ML) found by the program, which is stable with high Si-rich chemical potential.